\documentclass{bmcart}

\usepackage{amsthm,amsmath}
\usepackage[utf8]{inputenc} 
\usepackage{graphicx}
\usepackage{float}
\usepackage{amssymb}
\usepackage{caption, booktabs, tabularx, makecell}
\usepackage{longtable}
\usepackage{xspace}
\usepackage{multirow}
\usepackage{euflag}

\startlocaldefs
\endlocaldefs

\begin{document}

\begin{frontmatter}

\begin{fmbox}
\dochead{Survey}


\title{Cyber-Security Challenges in Aviation Industry: A~Review of Current and Future Trends}



\author[
  addressref={aff1},                   
  email={eaukwandu@cardiffmet.ac.uk}   
]{\inits{E.U.}\fnm{Elochukwu} \snm{Ukwandu}}
\author[
  addressref={aff2},                   
  email={mohamed.ben-farah@strath.ac.uk}   
]{\inits{M.A.B.F.}\fnm{Mohamed Amine} \snm{Ben Farah}}
\author[
  addressref={aff3},                   
  email={hananhindy@ieee.org}   
]{\inits{H.H.}\fnm{Hanan} \snm{Hindy}}
\author[
  addressref={aff4},                   
  email={buresm3@fel.cvut.cz}   
]{\inits{M. B.}\fnm{Miroslav} \snm{Bures}}
\author[
  addressref={aff2},                   
  email={robert.atkinson@strath.ac.uk}   
]{\inits{R. A.}\fnm{Robert} \snm{Atkinson}}
\author[
  addressref={aff2},                   
  email={christos.tachtatzis@strath.ac.uk}   
]{\inits{C.T.}\fnm{Christos} \snm{Tachtatzis}}
\author[
  addressref={aff2},  
  corref={aff2},
  email={xavier.bellekens@strath.ac.uk}   
]{\inits{X.B.}\fnm{Xavier} \snm{Bellekens}}

\address[id=aff1]{
  \orgdiv{Department of Computer Science},             
  \orgname{Cardiff School of Technologies, Cardiff Metropolitan University},          
  \city{Cardiff, Wales},                              
  \cny{UK}                                    
}
\address[id=aff2]{
  \orgdiv{Department of Electronic and Electrical Engineering},             
  \orgname{University of Strathclyde},          
  \city{Glasgow, Scotland},                              
  \cny{UK}                                    
}
\address[id=aff3]{
  \orgdiv{Division of Cyber-Security},             
  \orgname{Abertay University},          
  \city{Dundee, Scotland},                              
  \cny{UK}                                    
}
\address[id=aff4]{
  \orgdiv{Department of Computer Science},             
  \orgname{Faculty of Electrical Engineering, Czech Technical University in Prague},          
  \city{Prague},                              
  \cny{Czechia}                                    
}



\end{fmbox}


\begin{abstractbox}
\begin{abstract} 
The integration of Information and Communication Technology (ICT) tools into mechanical devices found in aviation industry has raised security concerns. The more integrated the system, the more vulnerable due to the inherent vulnerabilities found in ICT tools and software that drives the system. The security concerns have become more heightened as the concept of electronic-enabled aircraft and smart airports get refined and implemented underway. In line with the above, this paper undertakes a review of cyber-security incidence in the aviation sector over the last 20 years. The essence is to understand the common threat actors, their motivations, the type of attacks, aviation infrastructure that is commonly attacked and then match these so as to provide insight on the current state of the cyber-security in the aviation sector. The review showed that the industry's threats come mainly from Advance Persistent Threat (APT) groups that work in collaboration with some state actors to steal intellectual property and intelligence, in order to advance their domestic aerospace capabilities as well as possibly monitor, infiltrate and subvert other nations’ capabilities. The segment of the aviation industry commonly attacked is the Information Technology infrastructure, and the prominent type of attacks is malicious hacking activities that aim at gaining unauthorised access using known malicious password cracking techniques such as Brute force attacks, Dictionary attacks and so on. The review further analysed the different attack surfaces that exist in aviation industry, threat dynamics, and use these dynamics to predict future trends of cyber-attacks in the industry. The aim is to provide information for the cyber-security professionals and aviation stakeholders for proactive actions in protecting these critical infrastructures against cyber-incidence for an optimal customer service oriented industry.
\end{abstract}




\end{abstractbox}
%

\end{frontmatter}



\section{Introduction}
\label{Section: Introduction}
The critical issues of cyber-security have attracted much attention in the aviation industry, since the emergent of current efforts at integrating Information and Communication Technology (ICT) tools into mechanical devices found in aviation industry. Thus, forming part of the ongoing efforts of making the aviation industry compliant to the emerging 4th industrial revolution through smart airports and e-enabled aircraft projects~\cite{duchamp2016cyber}.\\

Because of the strategic position the aviation industry plays as well as being the gateway to other nations and the sensitive nature of the system, mistakes are deemed very costly. Minor errors or oversights may lead to fatality, loss or exposure of stakeholders', staff and customers' personally identifiable information, credentials and intellectual properties and intelligence theft. As major threat actors in the industry are found to be working in collaboration with state actors with the aim to steal intellectual property and intelligence, in order to advance their domestic aerospace capabilities as well as possibly monitor, infiltrate and subvert other nations’ capabilities. Thus commensurable cyber-defense strategies become imperative.\\

Monteagudo~\cite{Monteagudo2020aviation} suggests that industry players should employ micro-segmentation strategies in cyber-defense design and implementation. As micro-segmentation enables aviation network infrastructure to be divided into multiple micro-segments and to apply separate access privileges. The approach opined, helps to contain any compromise or data breach to its specific segment. Others such as Bellekens~\textit{et al.}\cite{bellekens2019cyber} proposed a deception solution for ensuring early detection of breaches in the aerospace and other critical infrastructure sectors. \\

In this manuscript, we explore the cyber-security situation in civil aviation industry. The term civil aviation is used to describe a category of flight operations that are non-military in nature, covering both private and commercial areas of the industry. This includes all parts of the aviation ecosystem, which also extend to the whole system of avionics, air-traffic controls, airlines, and airports. The essence is to take critical review of the current trends and using same to predict future trends in the industry with the introduction of modern Information Technology (IT) tools; such as Internet of Things (IoTs) devices, machine learning, cloud storage and cloud computing in aviation industry. \\

In the rest of the paper, Section~\ref{Section: A Systematic Literature Review} explored available literature on cyber-threats in civil aviation industry, the threat actors and their motivations. Section~\ref{Section: Documented Cyber-Attacks in Aviation Industry (2000-2020)} focuses on the documented cyber-attacks in civil aviation in the last 20 years. Section~\ref{Section: Feasible Attack Surfaces and Vulnerabilities in Aviation Industry} provides feasible attack surfaces a malicious attacker can exploit at the airport or aircraft systems and ways they can be mitigated. Section~\ref{Section: Civil Aviation} contains insight into types of cyber-attacks and reasons for such attacks in private and commercial flight operations and airports. Section~\ref{Section: Future of Civil Aviation Industry and their cyber-security challenges} gives the future of civil aviation as it relates to smart airports and e-enabled aircraft. While Section~\ref{Section: ThreatDynamicsandAnalyses} is about threat dynamics and their implications on the future of civil aviation industry. The paper concluded in Section~\ref{Section: Conclusion}.

\section{A Systematic Literature Review}
\label{Section: A Systematic Literature Review}
This section explores available literature on cyber-threats in civil aviation industry, the threat actors and what motivates them.

\subsection{\textbf{Cyber threats in civil aviation industry}}
The use and reliance on cyber-technologies have become an integral part of the aviation ecosystem, which also extend to the whole system of avionics, air-traffic controls, airlines, and airports~\cite{haass2016aviation}~\cite{Monteagudo2020aviation}. The impacts range from improving on the ground, air-borne or in-space operations, customer services; such as but not limited to ticket bookings, in-flight entertainments system, flights checking in and out, security screening of passengers, and use of aircraft cabin wireless network internet services~\cite{haass2016aviation}~\cite{lykou2018implementing}. It is also of no doubt that these technologies have been of great positive impacts to the aviation control systems, provided better aviation operations, safety, and performance~\cite{lykou2019smart}~\cite{Monteagudo2020aviation}~\cite{gopalakrishnan2013cyber}\cite{lykou2018implementing}~\cite{mathew2019airport}. In the same vein, the negative impacts have, in no measure, been quite devastating~\cite{Monteagudo2020aviation}~\cite{mathew2019airport}~\cite{suciu2018cyber} and hence this section focuses on the review of previous literature in areas of cyber-security issues in civil aviation industry over the last 20 years.\\

In 2018, Corretjer~\cite{corretjer2018cybersecurity} undertook a research to analyse the available cyber-security practices within the United States aviation industry (civil and military). The author also investigated efforts of the US government and private entities to protect the industry against cyber-attacks. The research summarised its findings by suggesting that although government’s Federal Airport Authority (FAA) and the private sector are putting in great efforts to tame the tides of cyber-attacks, more efforts are needed with regards to providing proactive measures against this menace during design, acquisition, operations and maintenance of aviation navigation systems.\\

With the introduction of modern IT tools such as IoT devices, machine learning, cloud storage and cloud computing in aviation industry, Kagalwalla and Churi~\cite{kagalwalla2019cybersecurity} are of the view that much attentions are needed in aviation cyber-security due to their inherent vulnerabilities. In the same vein, Duchamp, Bayram and Korhani~\cite{duchamp2016cyber} agree but added that the increase in the number of travellers, building of new modern airports, and complexities in new aircraft have brought with them an increase in cyber-attacks in civil aviation industry. \\

On the other hand, Lehto~\cite{lehto2020cyber} stated that the advancements in cyber-attack tools and methods as well as the increased exposures and motivation of the attackers have led to the current trend in these cyber-attacks thus affecting airlines, aircraft manufacturers and authorities. Cyber Risk International~\cite{CyberRisk2020cyber} submits that the rise in cyber-security challenges in the aviation industry is a result of the combination of digital transformation, connectivity, segmentation, and complexity currently being experienced in the industry due to surge in global travels. This further makes the industry to rely heavily on IT facilities to keep up with the pace, needs and transformations, thus getting it exposed to barrage of cyber-attacks. In all, they are of the opinion that with multiple entry and exit in aviation industry, creating a watertight defense is becoming a herculean task. Moreover, having lots of Legacy IT issues and fragmentation in the industry have in no measure increased the complexities as much of the IT systems in use were not designed to cope with modern challenges of cyber-crime \cite{Monteagudo2020aviation}. \\

Kagalwalla and Churi~\cite{kagalwalla2019cybersecurity} went further to say that in terms of securing the aviation industry against cyber-attacks factors like lack of resources, funds, and skilled staff have become part of the challenges. All the same, the issues of insider threats, procuring modern day operational technologies, like Supervisory Control and Data Acquisition (SCADA), Inter-Communication System (ICS), etc, remain part of it all. It finally offers solutions such as building strong security culture, implementing good preventing, and proactive measures as ways of confronting these challenges. \\

\subsection{\textbf{Threat actors and their motivations}}
\label{Subsection: ThreatActors}
The Fireeye Incorporated~\cite{Fireeye2016cyber} provided what they have observed as the major threat actors in aerospace industry and the motives behind these attacks. They went on to say that the industry cyber threats come mainly from Advance Persistent Threat (APT) groups that work in collaboration with state actors to steal intellectual property and intelligence, in order to advance their domestic aerospace capabilities as well as possibly monitor, infiltrate and subvert other nations’ capabilities. They also aim to develop countermeasures and produce technologies for sale on the global arms market they alleged. \\

They further support their claim, the authors provided information that through their threat intelligence have observed at least 24 APT compromise of organisation in different aerospace industry as well as the type of data stolen from the industry. This type of data range from budget information, business communications, equipment maintenance records and specifications. They also include Organisational Charts and Company Directories, Personally Identifiable Information, Product Designs, Product Blueprints, Production Processes and Proprietary Product or Service Information. Also included are, Research Reports, Safety Procedures, System Log Files and Testing Results and Reports.\\

Furthermore, Kessler and Craiger~\cite{kessler2018aviation} categorised the threat actors according to their motivations such as cyber-criminals, whose activities cost above 450 billion dollars annually to the global economy. While Cyber-activists/hacktivists, whose motives are mainly that of philosophy, politics, and non-monetary goals. Cyber-spies, on the other hand, are motivated by financial, industrial, political, and diplomatic espionage. Cyber-terrorists are driven by political, religious, ideological, or social violence. Finally, that Cyber-warriors are mainly attack by a nation-state in order to advance strategic goals. Abeyratne~\cite{abeyratne2020aviation}, added that according to adopted Resolution A40-10 at International Civil Aviation Organisation (ICAO) 40th General Assembly, threat actors have malicious intent focused in causing business disruptions, stealing information for political, as well as financial gains.\\

Based on the analysed literature, there are likelihood of rise in cyber-threats in civil aviation industry as global travel rate increases leading to reliance in IT tools to help keep up with the pace. Also, with embedded systems being deployed to help improve aviation services, a larger attack surface is provided due to the integration of hardware and software. Furthermore, with the threat actors having different motivational factors, it is therefore pertinent to state that there are high tendencies of increase in cyber-threats in civil aviation industry, alongside the dimension and type. This will depend on the actors involved, thus giving room for research and innovation to mitigate these risks, dissuade the actors by perhaps using proactive approaches of cyber-security by design in developing these IT tools.\\

\section{Documented Cyber-Attacks in Aviation Industry (2000-2020)}
\label{Section: Documented Cyber-Attacks in Aviation Industry (2000-2020)}
The reliance on technology, especially cyber-technology systems, have increasingly become part of the modern society. This reliance has in no doubt brought increased efficiency and effectiveness in day-to-day life, but it has also some attendant risks~\cite{Arampatzis2020the}. The reliance on cyber-enabled technologies have increased the safety and efficiency of air transport systems. In the same vein, a cyber-incident in one airport could pose a transnational problem with social and
economic consequences \cite{duchamp2016cyber}, due to high connectivity of human migration and the hyper-connectivity in aviation industry.\\

It is on the bases of the above that this section uses Tables~\ref{tab:CyberAttacks} to present reviews of documented cyber-threats in civil aviation industry over the last 20 years (2000-2020), as cyber incidents in aviation sector documented by for instance, Viveros in~\cite{viveros2016analysis} covered from 1997-2014.\\

\begin{center}

\begin{longtable}{|p{0.8cm}|p{0.8cm}|p{0.8cm}|p{2cm}|p{2cm}|p{7.5cm}|}

\caption{Cyber-Attacks in Civil Aviation Industry} \label{tab:CyberAttacks} \\

\hline \multicolumn{1}{|c|}{\textbf{Class}} & \multicolumn{1}{c|}{\textbf{Ref}} & \multicolumn{1}{c|}{\textbf{Year}}& \multicolumn{1}{c|}{\textbf{Incident}}& \multicolumn{1}{c|}{\textbf{Location}}& \multicolumn{1}{c|}{\textbf{Description}} \\ \hline 
\endfirsthead

\multicolumn{6}{c}%
{{\bfseries \tablename\ \thetable{} -- continued from previous page}} \\

\hline \multicolumn{1}{|c|}{\textbf{Class}} & \multicolumn{1}{c|}{\textbf{Ref}} & \multicolumn{1}{c|}{\textbf{Year}}& \multicolumn{1}{c|}{\textbf{Incident}}& \multicolumn{1}{c|}{\textbf{Location}}& \multicolumn{1}{c|}{\textbf{Description}} \\ \hline
\endhead

\hline \multicolumn{6}{|r|}{{Continued on next page}} \\ \hline
\endfoot

\hline
 \multicolumn{6}{|c|}{\textbf{Legend: C = Confidentiality, I = Integrity, A = Availability}}\\
\hline 
\endlastfoot

C&\cite{Gross2003faa}& 	2003&	Slammer Worm attack&	USA&	One of the FAA’s administrative server was compromised through a slammer worm attack. This attack shut down Internet service in some parts of Asia and slowed connections worldwide.\\
\hline
A&\cite{Goodin2009us}&2006&	Cyber-Attack&	Alaska, USA&	Two separate attacks on US Federal Aviation Administration (FAA) internet services that forced it to shut down some of its air traffic control systems.\\
\hline
C&\cite{Goodin2009us}&2008&Malicious hacking attack&	Oklahoma, USA&	Hackers stole administrative password of FAA’s interconnected networks when they took control of their system. By gaining access to the domain controller in the Western Pacific region, they were able to access more than 40,000 login credentials used to control part of the FAA's mission-support network.\\
\hline
C&\cite{Ellinor2009report}&2009&	Malicious hacking attack&	USA&	A malicious hacking attack on FAA’s computer, which gave them access to personal information on 48,000 current and former FAA employees.\\
\hline
C&\cite{Paganini2013Istanbul}&	2013&Malware attack & Istanbul, Turkey &Shutting down of passport control system at the departure terminals of Istanbul Ataturk and Sabiha Gokcen airports due to malware attack, leading to the delay of many flights.\\
\hline
C&\cite{Welshphishing}&	2013&	Hacking and Phishing attacks&	USA	&Malicious hacking and phishing attacks that targeted about 75 airports. These major cyber-attacks were alleged to have been carried out by an undisclosed nation-state sought to breach US commercial aviation networks.\\
\hline
A&\cite{Brewster2015attack}&	2015&	DDoS attack&	Poland&	A Distributed Denial of Service (DDoS) IT Network attack by cyber-criminals that affected LOT Polish Airlines flight-plan systems at the Warsaw Chopin airport. The attack made LOT’s system computers unable to send flight plans to the aircraft, thus grounding at least 10 flights, leaving about 1,400 passengers stranded.\\
\hline
I&\cite{Kirkliauskaite2020Main}&	2016&	Hacking, phishing
attacks&	Vietnam&	The defacement of website belonging to Vietnam airlines and flight information screens at Ho Chi Minh City and the capital, Hanoi, displaying messages of supportive China's maritime claims in the South China Sea by Pro-Beijing hackers.\\
\hline
A&\cite{Polityuk2016ukraine}&2016&Cyber-attack&Boryspil, Ukraine&A malware attack was detected in a computer in the IT network of Kiev’s main airport, which includes the airport’s air traffic control system.\\
\hline
A&\cite{Kirkliauskaite2020Main}&	2017&	Human error&
United Kingdom&	British flag-carrier computer systems failure caused by disconnecting and reconnection of the data-center power supply by a contracted engineer. This accident left about 75,000 passengers of British Airways stranded.\\
\hline
C&\cite{Park2018Cathay}&	2018&	Data breach&	Hong Kong&	Cathay Pacific Airways data breach of about 9.4 million customers’ personal identifiable information.\\
\hline
C&\cite{Sandlecyber}&	2018&	Data breach&	United Kingdom&	British Airways Data breach of about 380,000 Customers’ personal identifiable information.\\
\hline
C&\cite{Singh2018delta}&	2018&	Data breach&	USA	&Delta Air Lines Inc. and Sears Departmental stores reported a data breach of about 100, 000 customers’ payment information through third party.\\
\hline
A&\cite{Leyden2018brit}&	2018&	Ransomware attack&	Bristol Airport, UK&	An attack on electronic flight information screens at Bristol Airport. This resulted to the screen being taken offline and replaced with whiteboard information. There was no known adverse effect from this attack.\\
\hline
C&\cite{Sandle2018air}&	2018&	Mobile app data breach& Air Canada, Canada&	Air Canada reported a mobile app data breach affecting the personal data of 20,000 people.\\
\hline
C&\cite{Gibbs2018potential}&	2018&	Data breach&	Washington DC, USA&	Data breach on NASA server that led to possible compromise of stored personally identifiable information (PII) of employees on October, 23, 2018.\\
\hline
C&\cite{Gates2018boeing}&	2018&	Ransomware attack&	Chicago, USA&	Boeing was hit by the WannaCry computer virus. The attack was reported to have minimal damage to the company’s internal systems.\\
\hline
A&\cite{kessler2018aviation}&2018&Cyber-attack&Sweden&Cyber-attack launched by Russian APT group (APT28) that jammed Sweden's air traffic control capabilities, grounding hundreds of flights over a 5-day period.\\
\hline
A&\cite{Solomon2019israeli}& 	2019&	Bot attacks&	Ben Gurion Airport, Israel&	About 3 million bots attacks were blocked in a day by Israel’s airport authority as they attempted to breach airport systems.\\
\hline
C&	\cite{Duvelleroy2019airbus}&	2019&	Cyber-Incident&	Toulouse, France&	A cyber incident that resulted to an unauthorised access to Airbus “Commercial Aircraft business” information systems. There was no known impact according to the report on airbus’ commercial operations.\\
\hline
C&\cite{Goud2019ransomware}	&2019&	Ransomware attack&	Albany, USA&	Albany International Airport experienced a ransomware attack on Christmas of 2019. The attackers successfully encrypted the entire database of the airport forcing the authorities to pay a ransom in exchange of the decryption key to a threat actor.\\
\hline
C& \cite{Team2019cryptocurrency}&	2019&	Crypto mining Malware infection&	Europe&	A discovery through Cyberbit’s Endpoint Detection and Response (EDR) by Cyberbit researchers that showed an installation of crypto mining software infection that infected more than 50\% of the European airport workstations. \\
\hline
C&	\cite{Narendra2019privacy}&	2019&	Phishing attack&	New Zealand&	A phishing attack targeted at Air New Zealand Airpoints customers. This attack compromised the personal information of approximately 112,000 customers, with names, details and Airpoints numbers among the data exposed.\\
\hline
C&\cite{Montalbano2020doppelpaymer}&	2020&	Ransomware attack&	Denver, USA&	A cyber-incident that involved the attacker accessing and stealing company data. The stolen data were later leaked online.\\
\hline
C&\cite{Chua2020ransomware}&	2020&	Ransomware attack&	San Antonio, USA&	ST Engineering’s aerospace subsidiary in the USA suffered a data breach, which involved Maze Cyber-criminal gaining unauthorised access to its IT network and thus launched a ransomware attack.\\
\hline
I&\cite{Claburn2021Airline}&	2021&	Human Error &	Birmingham, United Kingdom & A flaw in the IT system used by the operator to produce the load sheet, meant that an incorrect takeoff weight was passed to the flight crew.\\
\hline
\end{longtable}

\end{center}

\subsection{\textbf{Analysis and Critical Reviews of Table \ref{tab:CyberAttacks}}}

\begin{figure}[hbt!]
    \centering
    \includegraphics[width=8cm]{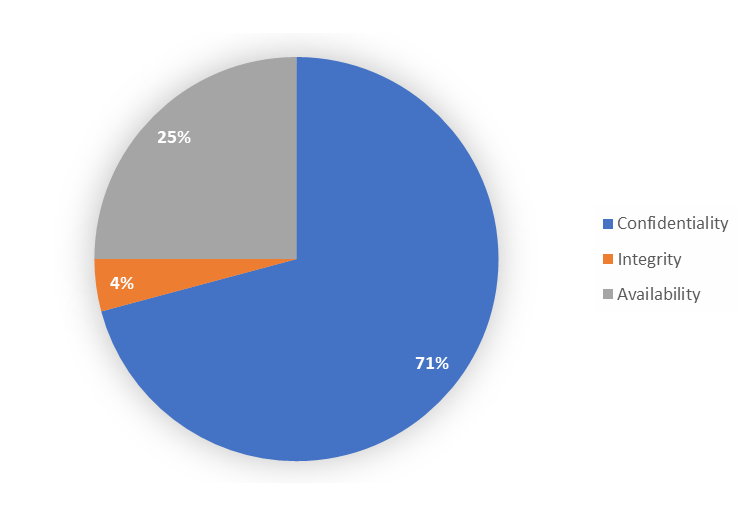}
    \caption{\textbf{Cyber-Attack Class based on Security Triad}}
    \label{fig:AttackClass}
\end{figure}

From Figure~\ref{fig:AttackClass}, attacks focusing on stealing login details such as administrative passwords, malicious hacking to gain unauthorised access in IT infrastructure are major focus of cyber-attackers in Aviation industry over the last 20 years with about 71\%. The second focuses on denial of service such as Distributed Denial of Services (DoS) to hinder data availability to prospective customers at 25\%, and the third are attacks that tends to undermine the integrity of files, by intercepting them while on transit or at rest so as to corrupt the contents with 4\%. These results provide credence to assertions in subsection~\ref{Subsection: ThreatActors}, which posits that the major motivation of threat actors is to steal intellectual properties and intelligence, in order to advance their domestic aerospace capabilities as well as possibly monitor, infiltrate and subvert other nations’ capabilities.\\

\begin{figure}[hbt!]
    \centering
    \includegraphics[width=8cm]{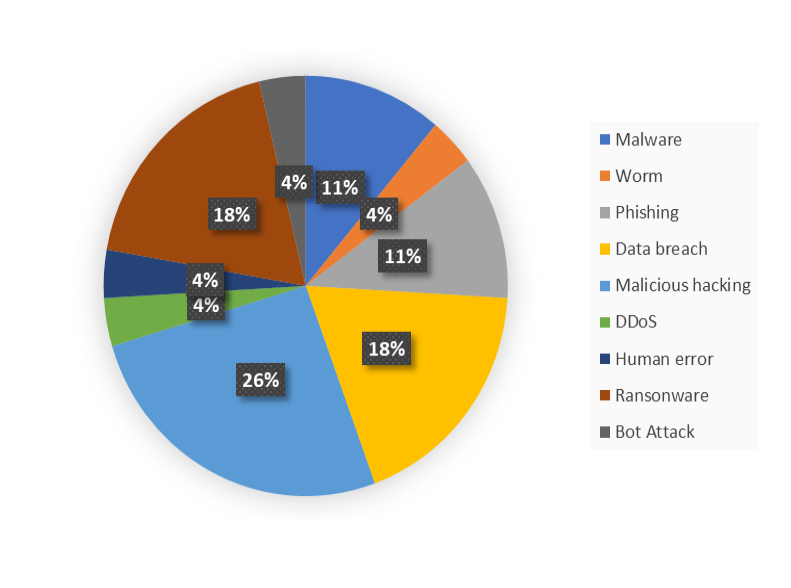}
    \caption{\textbf{Cyber-Attacks by Types}}
    \label{fig:AttackTypes}
\end{figure}

\begin{figure}[hbt!]
    \centering
    \includegraphics[width=8cm]{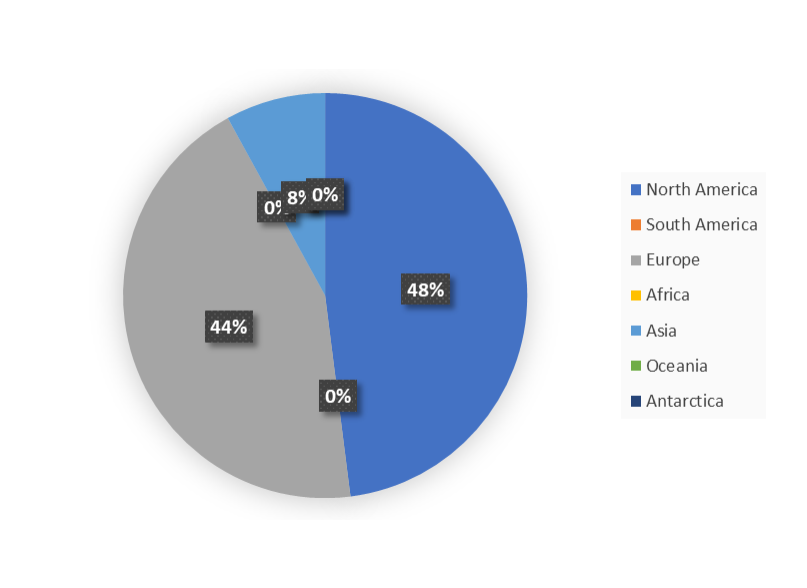}
    \caption{\textbf{Cyber-Attacks by Location}}
    \label{fig:AttackLocation}
\end{figure}

\begin{table*} [hb!]
\begin{center}
\caption{\textbf{Cost of Cyber-Attacks in Aviation Industry Per Year}}
\begin{tabular}{|l|c|c|c|c|}
\hline
Year&	No Of PersonsAffected&	MoneyLost&	Airports ShutDown&	Lost FlightHours\\
\hline
2003&	NA&	NA&	NA&	NA\\
\hline
2006&	NA&	NA&	2&	NA\\
\hline
2008	&40000&	NA&	NA&	NA\\
\hline
2009	&48000	&NA	&NA&	NA\\
\hline
2013	&NA	&NA&	77&	NA\\
\hline
2015	&1400	&NA	&NA&	NA\\
\hline
2016	&NA&	NA&	NA&	NA\\
\hline
2017	&75000	&NA	&NA	&NA\\
\hline
2018	&94500000&	NA&	NA&	120\\
\hline
2019	&112000	&NA	&NA	&NA\\
\hline
2020&	NA&	NA&	NA&	NA\\
\hline
\end{tabular}
\label{tab:AttackAnalyses}
\\Legend: NA = Record Not Available\\
\end{center}
\end{table*}

Results drawn from cyber attack by types in Figure~\ref{fig:AttackTypes} supports the evidence provided in Figure~\ref{fig:AttackClass}, by showing that malicious hacking activities top the list of types of cyber-attacks with 26\%, which tends to gain unauthorised access to IT infrastructure by breaking into it using known malicious password cracking techniques such as Brute force attacks, Dictionary attacks and so on. Data breach and Ransomware attacks are second with 14\% each, while attacks related to Phishing and Malware are third with 11\% each. Cyber incidents like Human error, Bot attacks, Worms and DDoS attacks are least with 4\% each. \\

Figure~\ref{fig:AttackLocation} results showed that cyber-attacks in Aviation industry are more in North America, with 11 out 12 recorded incidents coming from the United States of America (USA), and 1 only from Canada. This may not be unconnected with the large number of airports in the USA, as according to Mazareanu~\cite{Mazareanu2020number} in 2019, USA has about 5,080 public airports and about 14,556 private airports. Europe comes second with 44\% rate of the attack incidents with Britain topping the list of the countries in Europe frequently attacked. Countries in Asia continent come third with 8\% and least with countries in Africa with no known cyber-attacks recorded in their airports.\\

Table~\ref{tab:AttackAnalyses} results deal with the number of persons affected by cyber-incidents in aviation industry, monetary values lost either as compensations or as ransoms, number of times airports were shut down during cyber-attack incidents, and number of days air-crafts were grounded due to cyber-incidents at the airports. From the Table, 2018 remains top on the chart with highest rate of cyber-attacks in aviation industry, where about 94, 500, 000 persons were affected by cyber-attacks with about 5 days running of air-craft being grounded due to cyber-attack incidents at the airports. There was also a disturbing incident in 2019, as a result of Crypto mining Malware infection. This discovery was made by Cyberbit researchers through their Endpoint Detection and Response (EDR) software that showed an installation of crypto mining software infection that infected more than 50\% of the European airport workstations.\\

One major issue found with aviation cyber-security incidence is lack of transparency in record keeping, documentation and publication of these incidents for public knowledge. Take for instance, monies lost due to cyber-incidence were never publicised nor documented, especially the values paid by the industry as compensations to victims of these attacks. Other records not transparently disclosed are the number of shutdowns experienced by the various airports that were attacked as well as the lost flight hours during cyber-incidence.\\


\section{Cyber-Attack Surfaces and Vulnerabilities in Civil Aviation Industry}
\label{Section: Feasible Attack Surfaces and Vulnerabilities in Aviation Industry}
According to Paganini~\cite{paganini2014cyber}, only an attacker with a broad understanding on how an aircraft or aviation system works can successfully disrupt its normal operation. Thus, alluding to the fact that it is not an easy task to attack an entire aircraft or aviation system. on the other hand, Haass, Sampigethaya and Capezzuto~\cite{haass2016aviation}, highlighted that cyber-technologies like Wireless Fidelity (Wi-Fi), Internet protocols, IoT devices such as sensors, Global Positioning System (GPS), Open-source operating systems, Virtualisation, and Cloud computing services have assisted in aviation operations, making it cheaper, faster, and inter-operable. These systems, due to their different inherent vulnerabilities, can be targeted remotely by cyber-attackers, this position is also supported by~\cite{Thales2016overcoming}~\cite{Monteagudo2020aviation}, and Lykou~\textit{et al.}~\cite{lykou2018implementing}. All the same, Lykou~\textit{et al.}~\cite{lykou2018implementing} added that the practise of Bring Your Own Device (BYOD) by airport customers, travellers and employee have also constituted an attack surface to the industry. \\

Duchamp, Bayram and Korhani~\cite{duchamp2016cyber}, Kessler~\textit{et al.}~\cite{kessler2018aviation}, and Abeyratne~\cite{abeyratne2020aviation} are of the view that reliance in use of computer-based system in day-to-day management of aviation industry, which gave rise to improved sophistication in air navigation system, on-board aircraft control and communication system are some cyber-attack targets by malicious attackers. Also listed are, airport ground system, which includes flight information, security screening and day-to-day data management systems. It is based on these that this section presents some of the known feasible attack surfaces in Civil Aviation Industry (CAI), their known vulnerabilities and possible solutions.\\

In 2014, Santamarta~\cite{santamarta2014wake} discovered security flaws in Inmarsat and Iridium Satellite Communication (SATCOM) terminals, which are also in use in aviation industry. In the paper, the researcher concluded that malicious attackers have the potentials of exploiting the vulnerabilities inherent in the design of this system such as what appeared to be backdoors during their experiment. Also, exploitable are hardcoded credentials, insecure protocol, and weak encryption algorithms found in the system.\\

In 2017, Biesecker~\cite{biesecker2017boeing} reported that a demonstration by a team of government, industry and academic officials showed that a legacy Boeing 757 commercial plane was successfully hacked remotely in a non-laboratory setting by accessing the aircraft’s systems through radio frequency communications.\\

\subsection{\textbf{Aerospace and Avionic Systems}}
Aerospace systems embody much of software and hardware integration, being much of embedded-computing system technology. In view of this, the system is prone to software vulnerabilities as result of the combination as ensuring that embedded system is exempt from security flaws is a difficult one according to Dessiatnikof~\textit{et al.} in~\cite{dessiatnikoff2012potential} and Papp~\textit{et al.}~\cite{papp2015embedded}. The researchers in~\cite{dessiatnikoff2012potential} went further to assert from their findings that attacks on aerospace systems can originate from the lower layers such as Operating System (OS) kernel, protection mechanisms, and context switching as it is difficult even when formal verification methods are applied to prove absence of vulnerabilities in embedded systems. They concluded by saying that attacks against aerospace computer systems can be categorised based on the attacker's skills and aims. While one of the aims is usually to corrupt the computing system's core functions; the other on fault-tolerance mechanisms such as error detection and recovery systems.

On the other hand, aircraft avionics systems are critical to the safe operation of an airplane by crew members and pilots as it provides weather information, positioning data, and communications systems~\cite{GAO2020aviation}. Avionics is coined by combining aviation with electronics, which are made up of embedded systems used in aircraft design, development and operation~\cite{meyers1987encyclopedia}. Avionic systems through external sensors gather data such as speed, direction, and external temperature and route them to other components of aircraft using avionic network~\cite{smith2006system}.\\

In recent times, in a bid to leverage on low cost Commercial-Off-The-Shelf (COTS) equipment and software technology to increase bandwidth and reduce cost, Ethernet networks such as Avionics Full DupleX Switched Ethernet (AFDX) have been used in avionic network communication system. It is an IEEE 802.11 protocol-based Wireless Flight Management System (WFMS).\\

While this avionic communication network provide a secure wired network that is not easy for malicious users to access and inject false data, making it to provide a high degree of reliability and safety~\cite{akram2015challenges}~\cite{akram2016efficient}, Avionics Wireless Network (AWN) in the other hand, brings with it new challenges related to assurance, reliability and security~\cite{akram2016security}~\cite{markantonakis2016secure}. \\

Aircraft avionics provide passenger entertainment on-board an aircraft, but beyond this, they implement control of flight functions, navigation, control, guidance, communication, system operation, and monitoring through its software to hardware integrated systems. Because of this integration, some cyber-security concerns abounds, take for instance in Communications, where Voice over the Radio (VoR) was used to communicate with pilots and controllers. The major disadvantage of VoR is the time delay to receive the signal, especially in the case of multiple communications.
it is of no doubt that radio communication may give rise to several issues, such as the interruption of the signal or the misunderstanding between controller and pilot due to noise addition.
The solution is to use the Controller Pilot Data Link (CPDLC), which has the ability to send or receive the information as a digital signal.
The air carrier flight operations centres are synchronised with the flight deck to receive the same signal at the same time, allowing a maximum risk awareness and better decisions made.\\

Now, the aviation community is concentrated in creating modernised National Airspace System (NAS) and a new communication system that will be able to improve the interaction between the aircraft and the ground system.\\

More detailed attack surfaces on different aerospace and avionic components will be provided in the following sub-subsections.\\

\subsubsection{\textbf{Aircraft Communications Addressing and Reporting System (ACARS)}}
Aeronautical Radio, Incorporated (ARINC) introduced ACARS datalink protocol to help reduce crew workload and improve data integrity. ACARS is an ARINC 618-based air-to-ground protocols to transfer data between on-board avionics systems and ground-based ACARS networks~\cite{BellamyIII2018aviation}.\\

The ACARS system is made up of a Control Display Unit (CDU) and ACARS Management Unit (MU). While MU is to send and receive digital messages from the ground using existing very high frequency (VHF) radios. On the ground, the ACARS system (network of radio transceivers), receive (or transmit) the datalink messages, as well as route them to various airlines on the network.  \\

Smith~\textit{et al.} in~\cite{smith2017analyzing} and \cite{smith2018undermining} stated that current use of ACARS by stakeholders are beyond its original intention as conceived to serve as flight trackers and on crew automated timekeeping system. While in~\cite{smith2017analyzing} and~\cite{smith2018undermining} their works consisted of demonstrating how current ACARS usage systematically breaches location privacy, in~\cite{smith2017analyzing} showed how sensitive information sent with ACARS over a wireless channel can potentially lead to a privacy breach for users, supporting a known fact that ACARS message is susceptible to eavesdropping attack. While they concluded in~\cite{smith2017analyzing} by proposing a privacy framework, and in~\cite{smith2018undermining} recommended use of encryption and policy measures to tackle the known eavesdropping attack on the communication channel.\\

\subsubsection{\textbf{Automatic Dependent Surveillance-Broadcast (ADS-B)}}
Aircraft automatically transmits (ADS-B Out) and/or receives (ADS-B In) identification and positional data in a broadcast mode through a data link using Automatic Dependent Surveillance Broadcast (ADS-B). Through this means, the safety and capacity of airport surveillance are improved, thus enhancing situational awareness of airborne and ground surveillance in airports~\cite{ali2013safety}. Ali~\textit{et al.}~\cite{ali2017evaluation} in 2017 went further to support this claim by stating that ADS-B out provides varying ground applications support, which includes Air Traffic Control (ATC) surveillance in both radar and non-radar airspace on the airport surface. Thus, enabling enhanced surveillance applications by strengthening the capabilities of aircraft to receive ADS-B out message from other aircraft within their coverage (ADS-B In) areas. Sequel to this, the safety and credibility of ADS-B system is paramount as it plays its role in supporting various ground and airborne applications~\cite{ali2015safety}.\\

Furthermore, Manesh and Kaabouch in~\cite{manesh2017analysis} stated that in order to generate precise air picture for air traffic management, ADS-B employs global satellite navigation systems. As the system is designed to broadcast detailed information about aircraft, their positions, velocities, and other data over unencrypted data links, the security of ADS-B has become a major concern.\\

Tabassum~\cite{tabassum2017performance} analysed the performance of ADS-B data received from Grand Fork International Airport. The data format is in raw and archived Global Data Link (GDL-90) data format. GDL-90 is designed to transmit, receive and decode ADS-B messages via on-board datalink by combining GPS satellite navigation with datalink communications. The experiment was aimed at detecting anomalies in the data and quantifying the associated potential risk. In the course of the research, dropout, low confident data, message loss, data jump, and altitude discrepancy were identified as five different anomalies but the focus was on two - dropouts and altitude deviations. At the end of the analysis, the author concluded that all failures relating to the anomalies have potential of affecting ATC operation either from airspace perspective, such as Dropout, low confident data or from Aircraft perspective, such as Data jump, Partial message loss and Altitude discrepancy. In all these portend some level of attack surfaces, which an attacker can leverage on to carry out malicious intents such as Eavesdropping, Jamming attack, Message Injection, Message Deletion, and Message Modification~\cite{strohmeier2014security}~\cite{manesh2017analysis}.\\

\subsection{\textbf{Electronic Flight Bag}}
Electronic Flight Bag (EFB) is used to display digital documentation, such as navigational charts, operations manuals, and airplane checklists by the flight crew. It can also be used by the crew members to perform basic flight planning calculations. All the same, advanced EFB now exist for performing many complex flight-planning task. This are integrated into flight management systems alongside other avionic systems for use in displaying an airplane's position on navigational charts, with real-time weather information\cite{GAO2020aviation}.\\

Wolf, Minzlaff and Moser in~\cite{wolf2014information} stated that EFBs are attractive elements as they have replaced former paper references carried on board as part of the flight management system, thus bringing reduced weight to the aircraft system.  Currently, as advanced EFBs are integrated into flight management systems, unlike the previous ones that were stand-alone, they present an attack surface to the flight management system. Take for instance, a malware infected EFB could enable denial-of-service attacks to other connected on-board systems~\cite{GAO2020aviation}. This position, as stated above, is supported by~\cite{wolf2014information},~\cite{Howard2013Dell}, and~\cite{Keller2013Fokker} but they added that such infection is possible only with stand-alone EFBs.\\

\subsection{\textbf{Summarised Attack Surfaces in Civil Aviation Industry}}
Table~\ref{tab:AttackSurfaces} below contains available cyber-attack surfaces in civil aviation industry and our recommended ways to mitigate them.

\begin{center}

\begin{longtable}{|p{0.1cm}|p{1cm}|p{2cm}|p{3.5cm}|p{3.5cm}|p{3cm}|}

\caption{\textbf{Attack Surfaces and Components in Civil Aviation Industry}} \label{tab:AttackSurfaces} \\

\hline \multicolumn{1}{|c|}{\textbf{Class}} & \multicolumn{1}{c|}{\textbf{Ref}} & \multicolumn{1}{c|}{\textbf{Component}}& \multicolumn{1}{c|}{\textbf{Attack Surface}}& \multicolumn{1}{c|}{\textbf{Mitigation}}& \multicolumn{1}{c|}{\textbf{Description}} \\ \hline 
\endfirsthead

\multicolumn{6}{c}%
{{\bfseries \tablename\ \thetable{} -- continued from previous page}} \\

\hline \multicolumn{1}{|c|}{\textbf{Class}} & \multicolumn{1}{c|}{\textbf{Ref}} & \multicolumn{1}{c|}{\textbf{Component}}& \multicolumn{1}{c|}{\textbf{Attack Surface}}& \multicolumn{1}{c|}{\textbf{Mitigation}}& \multicolumn{1}{c|}{\textbf{Description}} \\ \hline
\endhead

\hline \multicolumn{6}{|r|}{{Continued on next page}} \\ \hline
\endfoot

\hline
 \multicolumn{6}{|c|}{\textbf{Legend: C = Confidentiality, I = Integrity, A = Availability}}\\
\hline 
\endlastfoot

C,I&\cite{santamarta2014wake}&SATCOM terminals&Hardcoded credentials, insecure protocol, weak encryption algorithms& Consistent patching and software updates, use of legacy encryption algorithm and network protocols.&SATCOM terminals can be exploited through some design flaws.\\
\hline
C,I&\cite{dessiatnikoff2012potential}\cite{papp2015embedded}&Aerospace systems&OS kernel, context switching, protection mechanisms&Consistent patching of OS, use of legacy encryption algorithm.&Attackers based on skills can exploit issues with integration of OS in embedded systems.\\
\hline
I&\cite{smith2017analyzing}\cite{smith2018undermining}&ACARS&Communication channel&Use of legacy encryption algorithm and policy measures. & ACARS communication channel is susceptible to eavesdropping and privacy breach.\\
\hline
I&\cite{tabassum2017performance}&ADS-B&Communication channel&Use of legacy encryption algorithm &ADS-B communication channel is prone to eavesdropping, Jamming attack, Message injection, deletion and Modification.\\
\hline
I&\cite{akram2016security}\cite{markantonakis2016secure}&AWN&Communication channel&Use of strong and legacy encryption algorithm&Wireless Avionic Network communication channel is prone to data integrity problems such as data assurance, reliability and security.\\
\hline
\end{longtable}

\end{center}

\section{Cyber-Security in Civil Aviation Industry}
\label{Section: Civil Aviation}
Civil aviation as stated earlier, includes private and commercial flight operations categorised into scheduled air transport and general aviation. Civil aviation industry plays very significant role in the global transportation and migration networks~\cite{taleqani2018machine}. Thus, the need to review the role of cyber-security in the industry, in relation to efforts being put in place to mitigate the attendant risks occasioned by cyber-attacks.\\

Cyber-attacks in aviation industry revolve around phishing and network attacks such as Eavesdropping, DoS, Man in the middle and spoofing attacks~\cite{taleqani2018machine}. Distributed Denial of Service (DDoS) and DoS attacks on network assets at the airport, especially Vulnerability Bandwidth Depletion DDoS Attacks (VBDDA) according to Ugwoke~\textit{et al.}~\cite{ugwoke2015security} could be mitigated by their proposed embedded Stateful Packet Inspection (SPI) based on OpenFlow Application Centric Infrastructure (OACI). Their focus was in using this technique to mitigate such attacks on the Airport Information Resource Management Systems (AIRMS). An enterprise cloud-based resource management system used in some airports. But Delain\textit{et al.}~\cite{Delain2016cyber} are of the position that DDoS could rather be prevented through Volumetric protection, providing an alternative secondary Internet connection, as well as implementing high performance hardware devices. The latter is to permanently monitor logging activities and traffic to improve the efficiency of the protection mechanism. \\

Clark and Hakim~\cite{clark2016cyber}, Martellini \cite{martellini2013cyber} and Singer and Friedman\cite{singer2014cybersecurity} posit the use of Airport intelligence classification to protect airport assets and infrastructure from cyber-attacks. This method they proposed has been classified as of good technical practice for high level security issues. The practice in real terms consist of good cyber-hygiene culture such as system and anti-virus regular updates, cyber-education for new employees, regular data backup and password management.\\

The use of encoding has been posited by Efe~\textit{et al.}~\cite{efe3air} as a measure to forestall cyber-attacks on ADS-B data used for airborne and ground surveillance in airports. They authors went further to say that the use of random blurring technique on aircraft data from ADS-B within a permissible error bounds for the purpose of Air Traffic Control (ATC), can be used to limit and monitor the level of interference of Unmanned Aerial Vehicles (UAVs) on ADS-B data using aircraft information at the airport.\\

\section{Future of Civil Aviation Industry and their cyber-security challenges}
\label{Section: Future of Civil Aviation Industry and their cyber-security challenges}
The concept of smartness in aviation industry is as a result of recent advancements in digitalisation efforts by integrating IoT enabled devices such as sensors in physical systems, use of Blockchain, AI, Cloud and Big Data technologies in service delivery. The essence is to provide optimal services, enhanced customer experience in a reliable and sustainable manner, and by working around the domains of growth, increase on efficiency, safety and security~\cite{lykou2019smart}. As increase in automation brings more attack surfaces due to increased IT integration into operation technologies, this section therefore looks at levels of cyber-security implementations, threats that evolved due to IoT and smart device integration, risk scenario analysis and possible mitigation as resilience measures.\\

\subsection{\textbf{Smart Airport}}
Smart Airport system is an integrated digital transformation within the airport ecosystem using new technologies such as IoT devices (sensors, actuators), Big Data, Cloud, and Blockchain technologies. Zamorano\textit{et al.} in~\cite{zamorano2020smart} included technologies like Code Bars Technology, Radio Frequency Identification (RFID), Geolocation technologies, Immersive Realities, Biometric Systems and Robotics as driving force in smart airport system. On the other hand, Koroniotis~\textit{et al.} in~\cite{koroniotis2020holistic} are of the view that advances in IoT device integration in aviation sector infrastructure have given rise to the emergence of smart airport. In all, its services are developed and processed to deliver good customer experience with improved efficiency in daily operation. It is also aimed at enhancing robustness, efficiency, and control in service delivery according to~\cite{koroniotis2020holistic}. This it does by gathering real-time customer data through interactions with every object at the airport and use same to analyse passenger’s profile and generate ancillary revenues~\cite{akarsmart}. In a nutshell, it is a data-rich environment, with equipment laced with range of sensors, actuators and other embedded devices, that provide customers a user-interface to interact with cyber-physical devices across the airport.\\

Georgia Lykou~\textit{et al.}~\cite{lykou2019smart} categorised threats against IoT applications in smart airports into Network and communication attacks, Malicious software and tampering with Airport smart devices. Others are misuse of authorisation; social engineering and phishing attacks. In all, the paper was focused on dealing with a complete scenario analysis of likely malicious attacks in smart airports with regards to IoT technologies, smart applications, mitigating actions, resilience measures and so on.\\

Koroniotis~\textit{et al.} in~\cite{koroniotis2020holistic} posits that IoT systems and networks due to likely hardware constraints, software flaws or IoT misconfigurations, IoT devices are prone to APT attacks. They suggest that the use of AI-enabled techniques such as Machine Learning can help in addressing the challenge of IoT-based cyber-attacks, and thus provide good cyber-defense to smart airport, strengthen reliability of services, and mitigate against service disruptions, travelling cancellations, or loss of sensitive information.

\subsection{\textbf{E-Enabled Aircraft}}
To make the aircraft more efficient, the use of electronic data exchange and digital network connectivity are paramount, according to Wolf~\textit{et al.}~\cite{wolf2014information} and in a bid to achieve this, IoT devices will play important role. This section reviews relevant works on the role of e-enabled devices in enhancing digital network connectivity and electronic data exchange in future e-enable aircraft. It also focuses on the potentials of these devices in shaping the future of aircraft industry with their attendant vulnerabilities, attack surfaces and possible mitigating factors.\\

Mahmoud~\textit{et al.} in 2010~\cite{mahmoud2010adaptive} undertook the designing of a proposed adaptive security architecture of future network connected aircraft system otherwise known as e-enabled aircraft. Other such works in related area were done by Neumann~\cite{neumann1997computer}, Sampigethaya~\textit{et al.}~\cite{sampigethaya2008secure}, Sampigethaya~\textit{et al.}~\cite{sampigethaya2011future}. While Mahmoud~\textit{et al.}~\cite{mahmoud2010adaptive} proposal is on a secure system topology for the embedded aircraft system network known as SecMan for application in Fiber-like aircraft Satellite Telecommunications. Sampigethaya~\textit{et al.} through this~\cite{sampigethaya2008secure} surveyed current and future security of embedded system in e-enabled aircraft network systems. In~\cite{sampigethaya2011future}, Sampigethaya~\textit{et al.} provided evidence that the safety, security, efficiency, etc of e-enabled aircraft will depend highly on the security capabilities of the communications, network and cyber-physical systems. \\

Because of the capabilities of e-enabled aircraft in having advance sensing, highly computerised systems, enhanced communication between on-ground and on-board systems, on-board system integration and some smart software-enabled interfaces, attack surface will likely increase. Such surfaces like exploiting internal cyber-physical system remotely through radio frequency jamming, node impersonation, passive eavesdropping, etc~\cite{sampigethaya2011future}. \\

There is no doubt that the integration of IT services into aircraft mechanical devices will improve efficiency in service delivery. All the same, it will also increase the attack surfaces and with the recent application of artificial intelligence techniques by cyber-attackers in automating their attack processes~\cite{kaloudi2020ai},~\cite{brundage2018malicious}, there is therefore need to work towards the use of AI-enabled cyber-defense strategies in the future e-enabled aircraft.

\section{\textbf{Threat Dynamics and Analysis}}
\label{Section: ThreatDynamicsandAnalyses}
With the increasing integration of IT to Operational Technology (OT) with the aid of Artificial Intelligence (AI) techniques by relying on data collected through sensors, thus giving OTs the ability to learn and operate in semi-autonomous or full autonomous states, this section aims to provide through available literature cyber-threat dynamics and their related analysis. While the dynamics provides recent advances in the automation of cyber-attacks using AI and Bio-inspired systems, the analysis is to determine the potential of threats, weaknesses, and vulnerabilities that can be exploited to achieve malicious goals using these modern attack technologies. In order to strengthen its focus, further attack classifications were provided using Table~\ref{tab:AttackClassed}. This section therefore provides an understanding of the possible threats and their characteristics, so as to inform relevant stakeholders on the optimum prevention and mitigation measures. 

\subsection{\textbf{Threat Dynamics}}
\subsubsection{\textbf{AI-based Attacks}}
The recent advances in AI have been embraced by cyber-criminals to automate attack processes~\cite{kaloudi2020ai},~\cite{brundage2018malicious}, taking advantage of technologically enhanced learning and automation capabilities offered by deep and reinforcement learning. The trend has necessitated the pressing need to develop appropriate cyber-situational awareness, cyber-hygiene, training methods, scenarios and technologies in response.\\

Kaloudi and Li~\cite{kaloudi2020ai} reported a list of existing AI-enhanced cyber-attacks; (1)~Next Generational Malware such as DeepLocker~\cite{kirat2018deeplocker} and Smart Malware~\cite{cohen1999simulating}. (2)~Voice Synthesis such as Stealthy Spyware~\cite{zhang2018using}. (3)~Password-based Attacks such as Next-generation password brute-force attack~\cite{trieu2018artificial} and PassGAN~\cite{hitaj2019passgan}. (4)~Social Bots such as: SNAP\textunderscore~R~\cite{seymour2016weaponizing}, DeepPhish~\cite{bahnsen2018deepphish} and Fake reviews attack~\cite{yao2017automated}. (5)~Adversarial Training such as MalGAN~\cite{hu2017generating}, DeepDGA~\cite{anderson2016deepdga} and DeepHack. The majority of these attacks targeted interconnected and software dependent new generational embedded systems known as Smart Cyber Physical Systems such as smart traffic management systems, smart healthcare systems, smart grids, smart buildings, autonomous automotive systems, autonomous ships, robots, smart homes and intelligent transport systems.

\subsubsection{\textbf{Bio-Inspired Attacks}}
The Backtracking Search Optimisation Algorithm~(BSA) and Particle Swarm Optimisation~(PSO) are two Active System Identification attacks developed by~\cite{de2017bio} using bio-inspired meta-heuristics~\cite{farah2019image} and tested in a controlled environment. The goal was to highlight the potential impacts of automated attacks, especially their degree of accuracy in damaging the Network Controlled Systems, as a stimulus to develop solutions that counter this attack class.
Chen~\textit{et al.}~\cite{chen2016decapitation} coined the term `A Bio-inspired Transmissive Attack', a scenario exemplified in Stuxnet~\cite{langner2011stuxnet},~\cite{farwell2011stuxnet},~\cite{chen2011lessons},~\cite{lindsay2013stuxnet}, best described as a stealthy breach utilising a biological epidemic model in the communication system to propagate the attack. In addition to the hidden nature of the attack, the hacker need not be conversant with the network topology to succeed. Hence, the linkage between transmissive attacks and epidemic models.

\subsection{\textbf{Threat Analysis}}
The essence of threat analysis is to determine the potential threats, weaknesses, and vulnerabilities that can be exploited to achieve malicious goals~\cite{stango2009threat}. An understanding of the possible threats and their characteristics informs on the optimum prevention, and mitigation measures. The optimum response is also governed by the existing risk mitigation policies for a specific architecture, functionality, and configuration as defined by regulating bodies. One of the challenging requirements is the metrics to be used to determine the status of the network security performance, the basis to define approaches to increase its robustness. \\

Table~\ref{tab:AttackClassed} provides some tabulated analysis on cyber-attacks by classifying them into domain of attacks, experimental tests, scenarios and tools for analysis. 

\subsubsection{\textbf{Modelling-based Approach}}
The aim is to predict the behaviour of unknown attacks and to create models able to prevent threats.  The actual vulnerability and security default of the system is core in order to conceptualise such a model. The configuration and architecture of the local network is a requirement in the development of a cyber-threat detection model.\\

Ibrahim~\textit{et al.}~\cite{ibrahim2014modelling} proposed the use of a formal logic known as Secure Temporal Logic of Action [S-TLA.sup+] as a modelling-based approach for reconstructing evidence of Voice Over Internet Protocol~(VoIP) malicious attacks. The goal of the research was to generate related additional evidence and to measure the consistency against existing approaches using the [S-TLA.sup+] model checker.\\

Mace~\textit{et al.}~\cite{mace2018multi} reported on a multi-modelling-based approach to assessing the security of smart buildings. The approach was based on an Integrated Tool Chain for Model-based Design of Cyber-Physical Systems~(INTO-CPS), a suite of modelling, simulation, and analysis tools for designing cyber-physical systems. The study was motivated by the evolution to smart buildings controlled by multiple systems that provide critical services such as heating, ventilation, lighting, and access control, all highly susceptible to cyber-attacks. The stages of a systemic methodology to assessing the security when subjected to Man-in-the-Middle attacks on the data connections between system components by using a fan coil unit case study was presented.\\

\begingroup
\scriptsize
\begin{longtable}{|m{0.1\linewidth}|>{
\centering\arraybackslash}m{0.05\linewidth} |m{0.4\linewidth}|m{0.4\linewidth}|}
\caption{\textbf{Attack Classifications}} \label{tab:AttackClassed} \\

\hline 
Domain &Ref & Experimental tests / Scenarios & Tools\\
\endfirsthead

\hline 
Domain &Ref & Experimental tests / Scenarios & Tools\\
\endhead

\hline
\multicolumn{4}{r}{{Continued \ldots}} \\
\endfoot

\endlastfoot

	
	
  \hline
	\multirow{7}{*}{IoT}& \cite{siboni2016security}&Network mapping attack/Implementation of profiling module (Training and testing algorithm) & TestStad/ Machine Learning Algorithm \\ \cline{2-4}
	
	&\cite{wang2019capacity}& Discrete-time Markov Chain model (DTMC): Analysing the capacity of the block chain& Block mining algorithm and Ethereum protocol \\ \cline{2-4}
														
	&\cite{waraga2020design}& Manual test: Analysis and attacks of each device, Automated test: process testing of different IoT device &Open Source MS \\ \cline{2-4}
    
    &\cite{lee2018design}& DoS massif trafic/Transfert Data/Abnormal code/System crash &DTM by Triangle Micro Works \\ \cline{2-4}
    
    &\cite{kim2019soda}& Real-world attack scenarios: internal and external network attacks &SDN/network function virtualisation \\ \cline{2-4}
	
	&\cite{shafiq2020selection}& Anomaly intrusion/ Attacks traffic &Machine Learning Algorithm/ Feature Extraction \\ \cline{2-4}
	
	&\cite{zolanvari2018effect}& Command injection attack &Machine Learning Algorithm/ PLC programming by Ladder language \\	 \cline{2-4}
	
	&\cite{elnour2020dual}&  SWaT/WADI datasets:Normal and attack scenario &Machine Learning Algorithm \\	 \cline{2-4}
	
	&\cite{molina2018enhancing} &  Man-in-the-middle attack  &SDN /Python \\  \cline{2-4}
	
	&\cite{arockia2019testbed}& LAUP algorithm(authentication)/ key distribution test &COOJA simulator \\
	\hline
	
	\multirow{4}{*}{Smart Grid}& \cite{hammad2019implementation} &Offline co-simulation Test-bed: DoS/FDI attacks& OMNET++ \\ \cline{2-4}
	
	& \cite{poudel2017real}& Access to communication link (\cite{hahn2013cyber}) attack model& OPAL-RT \\ \cline{2-4}
	&\cite{de2019implementation}& Deep packet inspection &Software Defined Networks/OpenFMB \\ \cline{2-4}
	
	&\cite{adepu2018epic}& Power supply interruption Attack/Physical damage attack &Real world power system/Machine learning \\		 \cline{2-4}	
	&\cite{fujdiak2019communication}& MMS/GOOSE/SV implementation &IEC 61850 Protocol/Ethernet RaspberryPi 3B+ \\	 \cline{2-4}
	
	&\cite{cheng2018development}&  HIL simulation/ proof-of-concept validation & Python \\ \cline{2-4}
	
	&\cite{liu2015integrated}&  DoS/Man in the middle attacks/TCP SYN Flood Attack &DeterLab/Security Experimentation EnviRonment (SEER) \\	 \cline{2-4}
	
	&\cite{oyewumi2019isaac}& Recording network traffic/Poisoning Attack &Real Time Digital Simulator (RTDS) \\	 \cline{2-4}
	
	&\cite{kezunovic2019testbed}& Timing Intrusion Attack &Field End-to-End Calibrator/ Gold PMU \\ \cline{2-4}
	
	&\cite{marino2019cyber}& Test of cyber-physical sensor: IREST &Idaho CPS SCADA Cybersecurity (ISAAC) testbed \\ \cline{2-4}

	&\cite{konstantinou2019flep}& MITM attack/DoS attack & Open source software/Raspberry Pis. FLEP-SGS \\	
	
		  \hline

	\multirow{4}{*}{Cloud}& \cite{patil2019designing}&Flood malicious traffic (ICMP/HTTP/SYN)& VMware Esxi hypervisor/A vCenter server/VMs \\ \cline{2-4}
	
	& \cite{celesti2019approach}& Considering  small messages (about1–2 KBytes):  Fast filling of the buffers & MOM4Cloud architectural model. \\ \cline{2-4}
															
	&\cite{mishra2020kvminspector}& UNM database: Malicious tracing logs &KVM2.6.27 hypervisor/ Python3.4 \\ \cline{2-4}
	
    &\cite{van2016performance}& Test of memory usage before/after instance creation & OpenStack: Open-Source cloud operating system \\ \cline{2-4}
    
    &\cite{ullah2019design}&  Evaluation of performance metrics of NDN/edge cloud computing & Cloud VM  \\ \cline{2-4}
    
    &\cite{al2018remote}& Adding defaults: broken interconnection/Abnormal extruder &MTComm: Online Machine Tool Communication  \\ \cline{2-4}
    
    &\cite{sanatinia2017hyperdrive}& Side channel attacks/   stealthy data exfiltration &DHCP server/TFTP Server/HTTP Server/MQTT Server  \\ \cline{2-4}
    
    &\cite{frank2017design}& SQL Injection attack &OpenStack implementation/Python  \\ \cline{2-4}
    
    &\cite{gao2015cyber}& Testing  traffic scenarios & Openflow controller/OpenvSwitch/Network virtualization agent \\ \cline{2-4}
	
	&\cite{khorsandroo2018time}& Time inference attacks &Software Defined Network \\ \cline{2-4}	
	
	&\cite{kalliola2017testbed}& DDoS attack &OpenStack environment \\
		  \hline

\end{longtable}
\endgroup

\section{Conclusion}
\label{Section: Conclusion}
With the emerging industrial revolution, the need for every sector of human endeavours to automate service delivery has become very prominent in both research, industry and service sectors. This paper focused on cyber-attack incidents in aviation industry for the last 20 years by first reviewing different records on cyber-attacks in civil aviation industry and the motives of the threat actors. From the review results of documented cyber-attacks in aviation industry within the years under review, the industry cyber threats as alleged come mainly from APT groups that work in collaboration with state actors to steal intellectual property and intelligence, in order to advance their domestic aerospace capabilities as well as possibly monitor, infiltrate and subvert other nations’ capabilities. \\

With the high demand for automation, the use of IoT devices has become prominent in order to engender high network connectivity between onground and onboard systems as well as provide Aircraft with advanced sensing capabilities. Furthermore, its integration in smart airport concept is aimed at delivering good customer experience with improved efficiency in daily operation. It is also aimed at enhancing robustness, efficiency, and control in service delivery. With this high-level integration and connectivity, cyber-attack surfaces are meant to increase as well. More worrisome is the ability of attackers to automate their attack processes using AI techniques, and hence this paper posits that to provide holistic cyber-defense strategies in the emerging Smart Airport and e-enable aircraft, the application of machine learning techniques in attack defense has become essential and exigent. This is because there are chances that APT group could advance beyond attacking only airport facilities to onboard and on the air Aircraft by using sophisticated remote attack tools.\\


\begin{backmatter}


\section*{Funding}
{\huge\euflag} The research is supported by the European Union Horizon 2020 Programme under Grant Agreement no. 833673. The content reflects the authors’ view only and the Agency is not responsible for any use that may be made of the information within the paper.




\section*{Competing interests}
The authors declare that they have no competing interests.

\section*{Consent for publication}
 All authors have read and agreed to the published version of the manuscript.

\section*{Authors' contributions}
This paper was conceptualised and formally analysed by Elochukwu Ukwandu, Mohamed Amine Ben-Farah, Hanan Hindy, and Xavier Bellekens.  While Elochukwu Ukwandu and Hanan Hindy carried out the investigation. The methodology was developed by Elochukwu Ukwandu, Xavier Bellekens and Hanan Hindy. Furthermore, the administration and supervision of this paper were carried out by Xavier Bellekens, while Miroslav Bures, Elochukwu Ukwandu, Richard Atkinson, Christos Tachtatzis and Xavier Bellekens did the validation. Formal writing of the paper that produced original draft were done by Elochukwu Ukwandu and Hanan Hindy. Miroslav Bures, Elochukwu Ukwandu, Hanan Hindy, Mohamed Amine Ben-Farah, Richard Atkinson, Christos Tachtatzis and Xavier Bellekens reviewed and edited the paper. All authors have read and agreed to the published version of the manuscript. 



\bibliographystyle{bmc-mathphys} 
\bibliography{Bibliography.bib}      


\end{backmatter}
\end{document}